\documentclass{PoS}

\usepackage{graphicx}
\usepackage{subfigure}
\usepackage{amssymb}
\usepackage{amsmath}
\usepackage{mathrsfs}
\usepackage{dsfont}
\usepackage{amsfonts}

\newcommand{\refeq}[1]{(\ref{#1})} 

\title{\centering Leading-order hadronic contribution to the anomalous magnetic moment of the muon from $N_f=2+1+1$ twisted mass fermions}

\ShortTitle{Leading hadronic contribution to $(g-2)_{\mu}$ from $N_f=2+1+1$ twisted mass fermions}

\author{Florian Burger \\
        Humboldt-Universit\"at zu Berlin, Institut f\"ur Physik, D-12489 Berlin, Germany\\
        E-mail: \email{florian.burger@physik.hu-berlin.de}}

\author{Xu Feng\\
        High Energy Accelerator Research Organization (KEK), Tsukuba 305-0801, Japan\\
        E-mail: \email{xufeng@post.kek.jp}}

\author{\speaker{Grit Hotzel} \\
        Humboldt-Universit\"at zu Berlin, Institut f\"ur Physik, D-12489 Berlin, Germany\\
        E-mail: \email{grit.hotzel@physik.hu-berlin.de}}

\author{Karl Jansen\\
        NIC, DESY, Plantanenallee 6, D-15738 Zeuthen, Germany\\
        E-mail: \email{karl.jansen@desy.de}}

\author{Marcus Petschlies\\
        The Cyprus Institute, P.O. Box 27456, 1645 Nicosia, Cyprus\\
        E-mail: \email{m.petschlies@cyi.ac.cy}}

\author{Dru B. Renner \thanks{Current address: Los Alamos National Laboratory}\\
        Jefferson Lab, 12000 Jefferson Avenue, Newport News, VA 23606, USA\\
        E-mail: \email{dru@jlab.org}}

\abstract{We present results for the leading order QCD correction to the anomalous magnetic moment of the muon including the first two
generations
of quarks as dynamical degrees of freedom. Several light quark masses are examined in order to yield a controlled extrapolation to the
physical pion mass. We analyse ensembles for three different lattice spacings and several volumes in order to investigate lattice artefacts
and finite-size effects, respectively. We also provide preliminary results for this quantity for two flavours of mass-degenerate
quarks at the physical value of the pion mass.}

\FullConference{31st International Symposium on Lattice Field Theory - LATTICE 2013\\
		July 29 - August 3, 2013\\
		Mainz, Germany}

\begin{document}

\section{Introduction}
\label{sec:introduction}
 
In search of physics beyond the standard model (SM) of elementary particle interactions, the anomalous magnetic moment of the muon,
$a_{\mu}$, is considered as one promising quantity which might disclose the nature of this new physics. There exists a
discrepancy between the experimental determination~\cite{Bennett:2006fi, Roberts:2010cj} of this observable and the SM calculation varying
between $2.4 \sigma$~\cite{Davier:2010nc} and $4.9 \sigma$~\cite{Benayoun:2012wc}. The big variation in the significance of the difference
arises from the determination of the hadronic contribution, $a_{\mu}^{\rm had}$, in the theoretical computation, since this cannot be
computed reliably in perturbation theory and therefore usually a dispersion relation is applied which relates $a_{\mu}^{\rm had}$ to other
experimental data. Hence, for the standard method employed so far the theoretical determination of this important quantity depends not only
on various model assumptions but also on the choice of experimental data.

In recent years, it has been shown that the lattice formulation of QCD constitutes a first-principle, ab-initio alternative to quantify at
least the leading-order hadronic contribution to $a_{\mu}^{\rm had}$ due to the insertion of the hadronic vacuum
polarisation~\cite{Blum:2002ii, Aubin:2006xv, Feng:2011zk, Boyle:2011hu}, $a_{\mu}^{\rm hvp}$, which currently is the
largest source of uncertainty in the SM calculation. These older computations, however, only took up to $N_f =2+1$ dynamical quark flavours
into account. Therefore, they could not unambiguously be compared with the results from the dispersive analyses because those as well as the
experimentally obtained values for $a_{\mu}$ at the current level of precision are sensitive to the complete first two generations of
quarks.

We remedy this shortcoming by performing computations on ensembles incorporating $N_f=2+1+1$ dynamical twisted-mass fermions generated
by the European Twisted Mass Collaboration (ETMC)~\cite{Baron:2010bv}. Preliminary results have been presented at Lattice
2012~\cite{Feng:2012gh}. Here, we additionally investigate systematic effects and for the first time attempt to perform the continuum
limit.
In particular, we check the validity of our chiral extrapolation
performed along the lines of Ref.~\cite{Feng:2011zk} for the light quark contribution, $a_{\mu, \rm ud}^{\rm hvp}$, by comparing the
extrapolated value at the physical point with the outcome of the calculation on an $N_f=2$ ensemble featuring the physical pion
mass~\cite{bartek:2013}.

\section{Basic Definitions}
The leading-order hadronic contribution to the muon anomalous magnetic moment can be computed directly in Euclidean
space-time~\cite{Blum:2002ii}
\begin{equation}
\label{eq:amudef}
a_{\mathrm{\mu}}^{\mathrm{hvp}} = \alpha^2 \int_0^{\infty} \frac{d Q^2 }{Q^2} w\left( \frac{Q^2}{m_{\mathrm{\mu}}^2}\right)
\Pi_{\mathrm{R}}(Q^2) 
\end{equation}
from the known function $w$ and the renormalised vacuum polarisation function
\begin{equation}
 \Pi_{\mathrm{R}}(Q^2)= \Pi(Q^2)- \Pi(0)\,.
\end{equation}
The vacuum polarisation function is related to the vacuum polarisation tensor by the requirement of transversity
\begin{equation}
   \Pi_{\mu \nu}(Q)= (Q_{\mu} Q_{\nu} - Q^2 \delta_{\mu\nu}) \Pi(Q^2)\,.
\end{equation}
Therefore, we first have to determine the vacuum polarisation tensor. It is given by the correlator of two electromagnetic vector currents.

Details on the lattice calculation as well as on the four-flavour twisted-mass ensembles analysed for this work can be found
in~\cite{Burger:2013jya}. Besides those $N_f=2+1+1$
ensembles we have also performed the calculation of the light quark contribution on a new $N_f=2$ ensemble possessing a pion mass very
close to its physical value. For details concerning this ensemble we refer to~\cite{bartek:2013}. 

Since the $N_f=2+1+1$ ensembles exhibit pion masses $m_{PS} \ge 227\,{\rm MeV}$, the question of how to extrapolate to the physical point
still remains important for our computation of $a_{\mu}^{\rm hvp}$. As mentioned before, we closely follow Ref.~\cite{Feng:2011zk} and use
a modified lattice definition
\begin{equation}
\label{eq:amudefresc}
 a_{\overline{\mathrm{\mu}}}^{\mathrm{hvp}} = \alpha^2 \int_0^{\infty} \frac{d Q^2 }{Q^2} w\left( \frac{Q^2}{H^2}
\frac{H_{\mathrm{phys}}^2}{m_{\mathrm{\mu}}^2}\right) \Pi_{\mathrm{R}}(Q^2) 
\end{equation}
which goes to $a_{\mathrm{\mu}}^{\mathrm{hvp}}$ for $m_{PS} \rightarrow m_{\pi}$, i.e.~when the hadronic scale $H$ determined at
unphysical pion masses, $m_{PS}$, attains its physical value $H_{\mathrm{phys}}$. This can also be viewed as a lattice redefinition of the
muon mass
\begin{equation}
 m_{\overline{\mu}} = m_{\mu}\cdot \frac{H}{H_{\rm phys}}\,.
\end{equation}
For the sake of consistency, we will always use $H=m_V$, the unphysical $\rho$-meson mass, even for the contributions originating from the
heavy quark currents.

As mentioned before, the target of the lattice computation is the determination of the vacuum polarisation function in order to use this as
input for the integrals in Eqs.~\refeq{eq:amudef} and~\refeq{eq:amudefresc}. Recently, there have been several suggestions for
fitting the vaccuum polarization function ~\cite{Aubin:2012me,Golterman:2013vna}.
Our way of performing the fits is
 described in detail in~\cite{Burger:2013jya}. Note that this differs from what is referred to as VMD+ (vector meson dominance plus a
 linear term in $Q^2$) in~\cite{Golterman:2013vna}.

\section{Results}
\subsection{The light quark contribution, $a_{\mathrm{\mu, ud}}^{\mathrm{hvp}}$}
\begin{figure}[htb]
\begin{minipage}{0.53\textwidth}
\includegraphics[width=\textwidth]{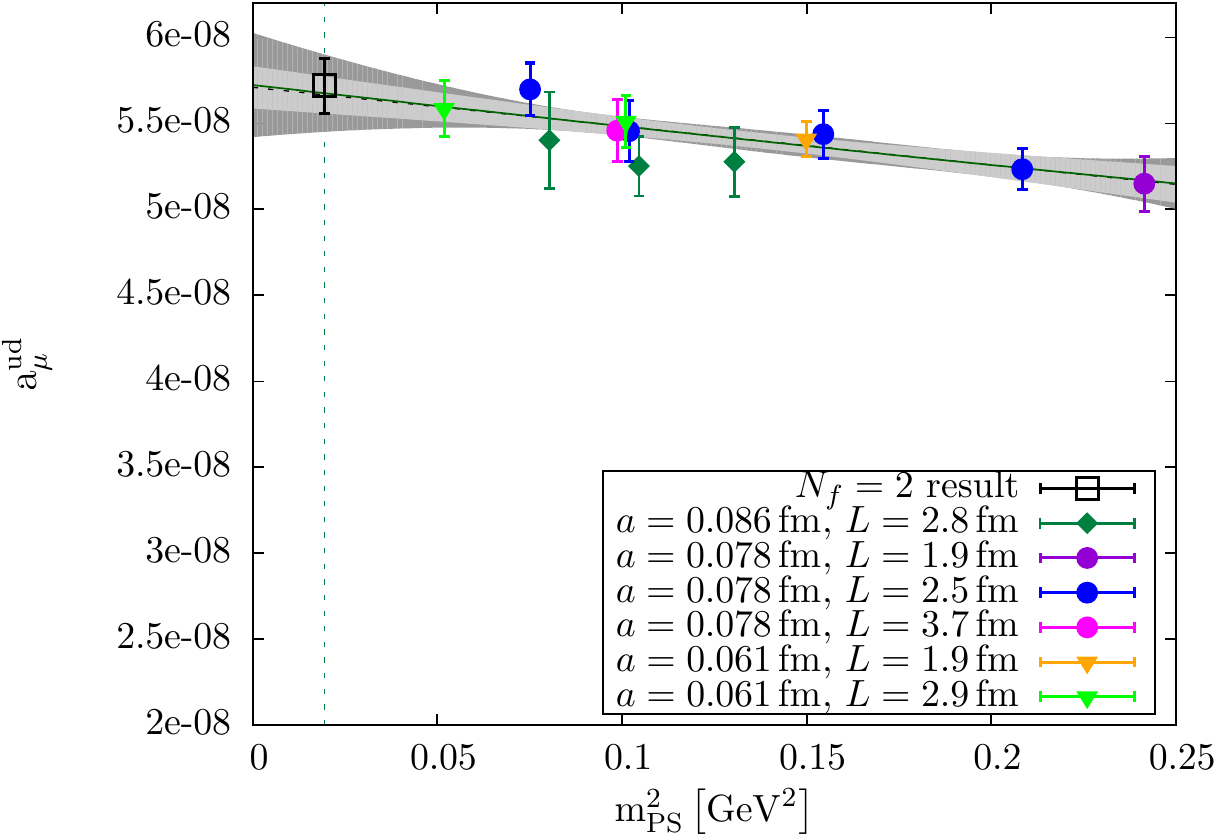}
\caption{Light-quark contribution to $a_{\mathrm{\mu}}^{\mathrm{hvp}}$ on $N_f=2+1+1$ sea.} 
\label{fig:amulight}
\end{minipage}
\hspace{0.02\textwidth}
\begin{minipage}{0.4\textwidth}
\includegraphics[width=\textwidth]{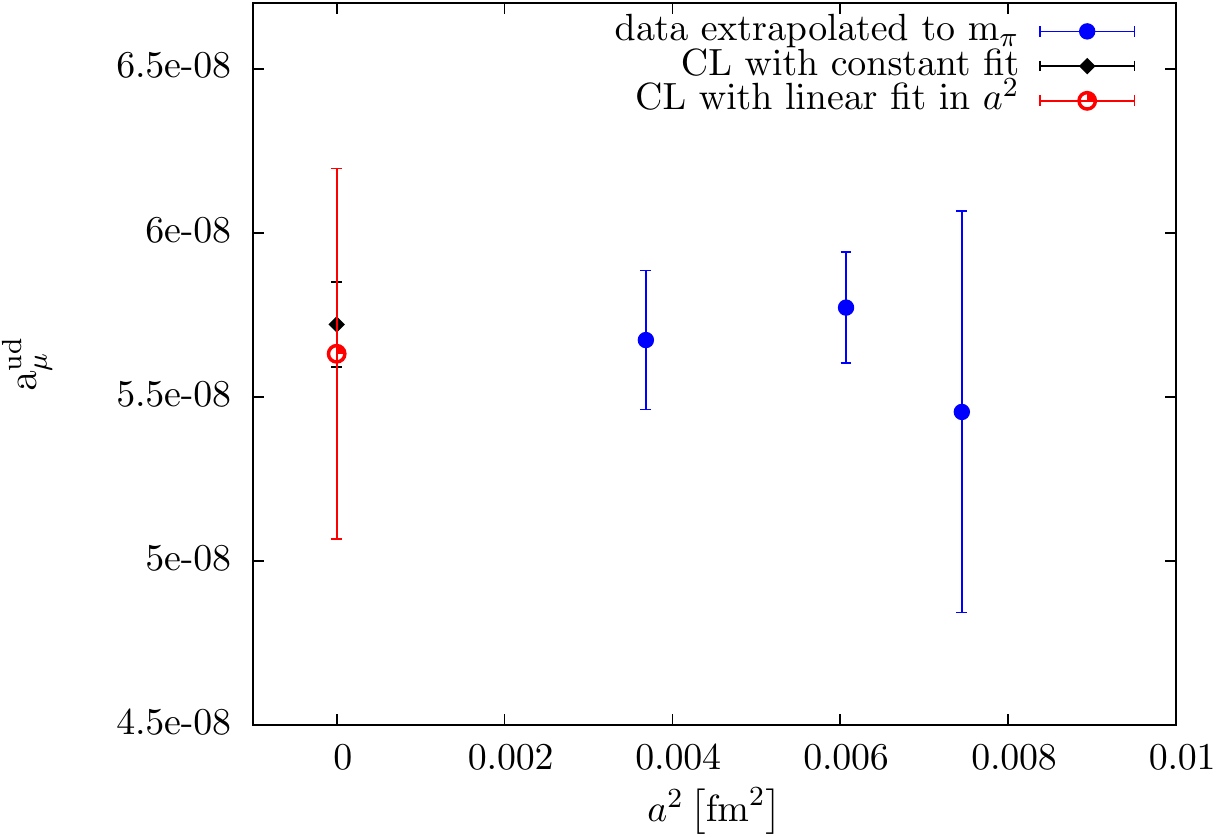}
\caption{Continuum extrapolation of $a_{\mathrm{\mu, ud}}^{\mathrm{hvp}}$.} 
\label{fig:amulight_cl} 
\end{minipage}
\end{figure}

The contribution of the valence up and down quarks to the total $a_{\mathrm{\mu}}^{\mathrm{hvp}}$ on the four-flavour sea is depicted in
Fig.~\ref{fig:amulight}. Here, we have employed the redefiniton of Eq.~\refeq{eq:amudefresc}, $a_{\mathrm{\overline{\mu},
ud}}^{\mathrm{hvp}}$, with $H=m_V$, the unphysical value of the $\rho$mass. The extrapolation to the physical point can be performed
by a simple
linear fit in the squared pseudo-scalar mass (black dotted line with light-grey error band). For comparison we also show the result of a
quadratic fit (dark-green solid line with dark-grey error band). A fit including an $a^2$-term gives a coefficient of this term which
is compatible with zero. This finding is also corroborated by first extrapolating linearly to the physical point for each individual lattice
spacing and then taking the limit $a \to 0$ as can be seen in Fig.~\ref{fig:amulight_cl}. Hence, at the current level of accuracy of our
data we cannot distinguish lattice artefacts in $a_{\mathrm{\mu, ud}}^{\mathrm{hvp}}$ and thus extrapolate the data from lattices of
different lattice spacings simultaneously.

Comparing the result of our linear fit with the one obtained on a two-flavour sea~\cite{Feng:2011zk}
\begin{eqnarray}
a_{\mathrm{\mu},\mathrm{ud}}^{\mathrm{hvp}} & = & 5.67(11)\cdot 10^{-8}\;\;\; (N_f=2+1+1)  \nonumber\\
a_{\mathrm{\mu},\mathrm{ud}}^{\mathrm{hvp}} & = & 5.72(16)\cdot 10^{-8}\;\;\; (N_f=2)\, ,
\label{eq: lighcontribution}
\end{eqnarray}
we find compatible results showing that the influence of the heavy sea quarks on the light valence quark contribution is small. This
together with the finding that we cannot discriminate lattice artefacts at the moment allows
us to check the validity of the redefinition and our chiral extrapolation by performing the calculation directly at the physical point  for
our new $N_f=2$ ensemble~\cite{bartek:2013}. 

\begin{figure}[htb]
 \centering
\includegraphics[width=0.53\textwidth]{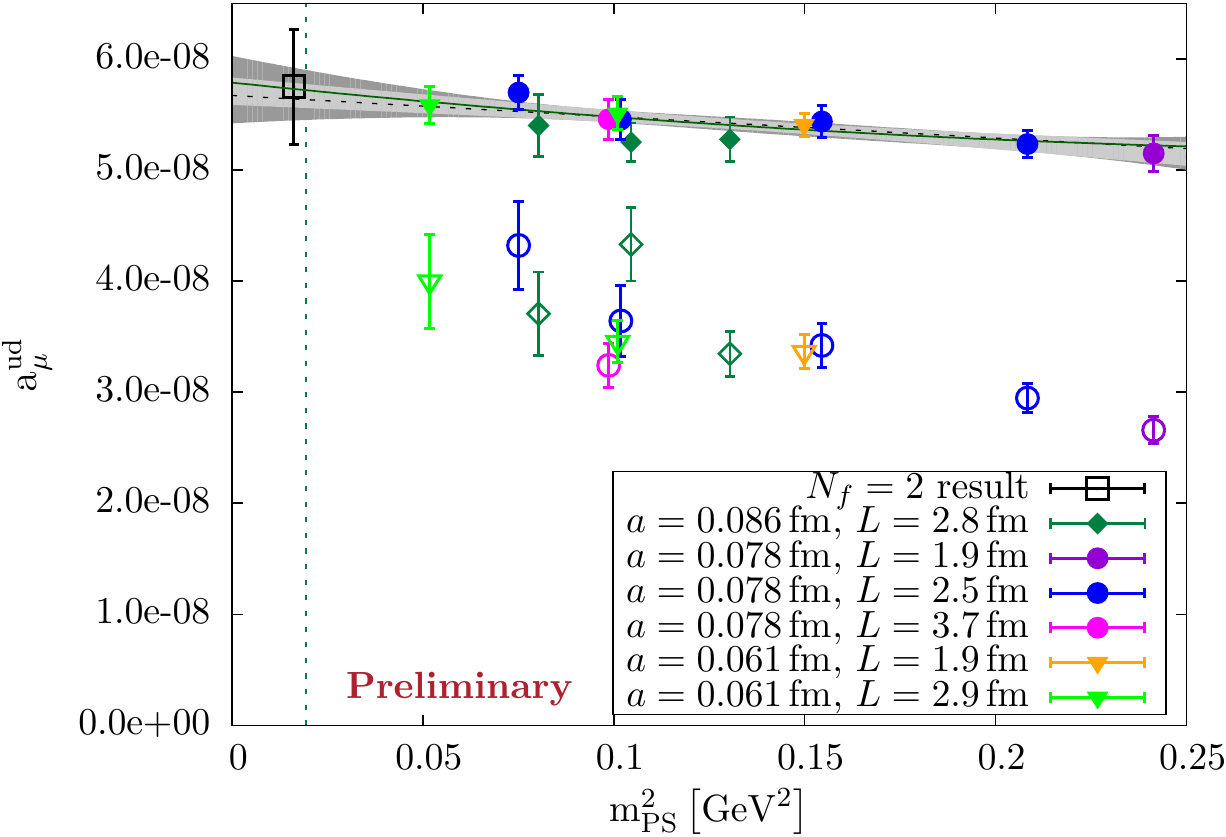}
\caption{Light-quark contribution to $a_{\rm \mu}^{\rm hvp}$ with filled symbols representing points obtained with
Eq.~\protect\refeq{eq:amudefresc} and open symbols obtained using Eq.~\protect\refeq{eq:amudef}. }
\label{fig:amulight_wphys}
\end{figure}

In Fig.~\ref{fig:amulight_wphys} besides the results already shown in Fig.~\ref{fig:amulight} also the values obtained with the standard
definition Eq.~\refeq{eq:amudef} are depicted as open symbols. In particular, the preliminary value at the physical point has been computed
employing Eq.~\refeq{eq:amudef} and agrees within its rather large uncertainty with the result obtained by linearly extrapolating the
two-flavour results on the four-flavour sea. This provides confidence in the validity of the redefinition in Eq.~\refeq{eq:amudefresc} and
the
related simple linear extrapolation in the squared pion mass.

\subsection{The three-flavour contribution, $a_{\mathrm{\mu, uds}}^{\rm hvp}$}
\begin{figure}[htb]
\begin{minipage}{0.4\textwidth}
\includegraphics[width=\textwidth]{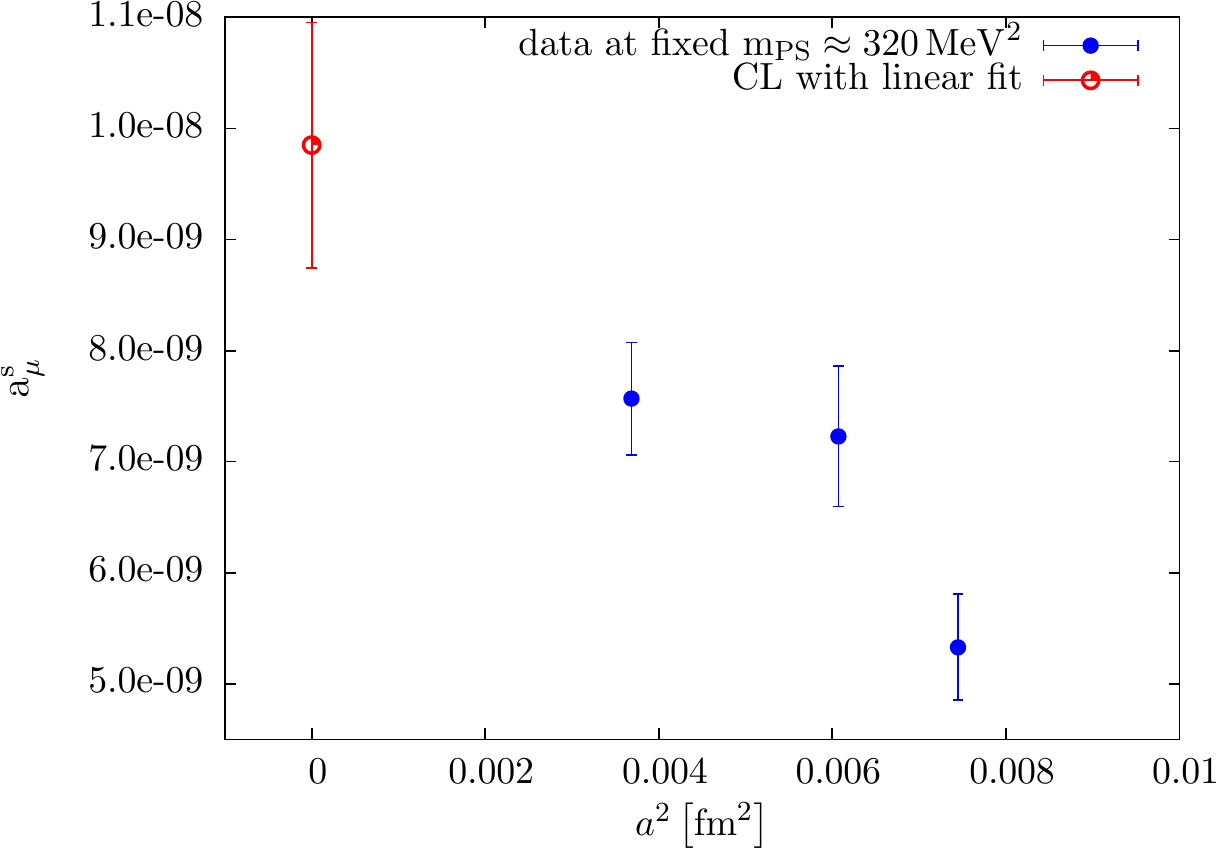}
\caption{Continuum extrapolation of $a_{\mathrm{\mu, s}}^{\rm hvp}$ at $m_{\rm PS}\approx320\,{\rm MeV}$.}
\label{fig:amustrange_cl}
\end{minipage}
\hspace{0.02\textwidth}
\begin{minipage}{0.53 \textwidth}
\includegraphics[width=\textwidth]{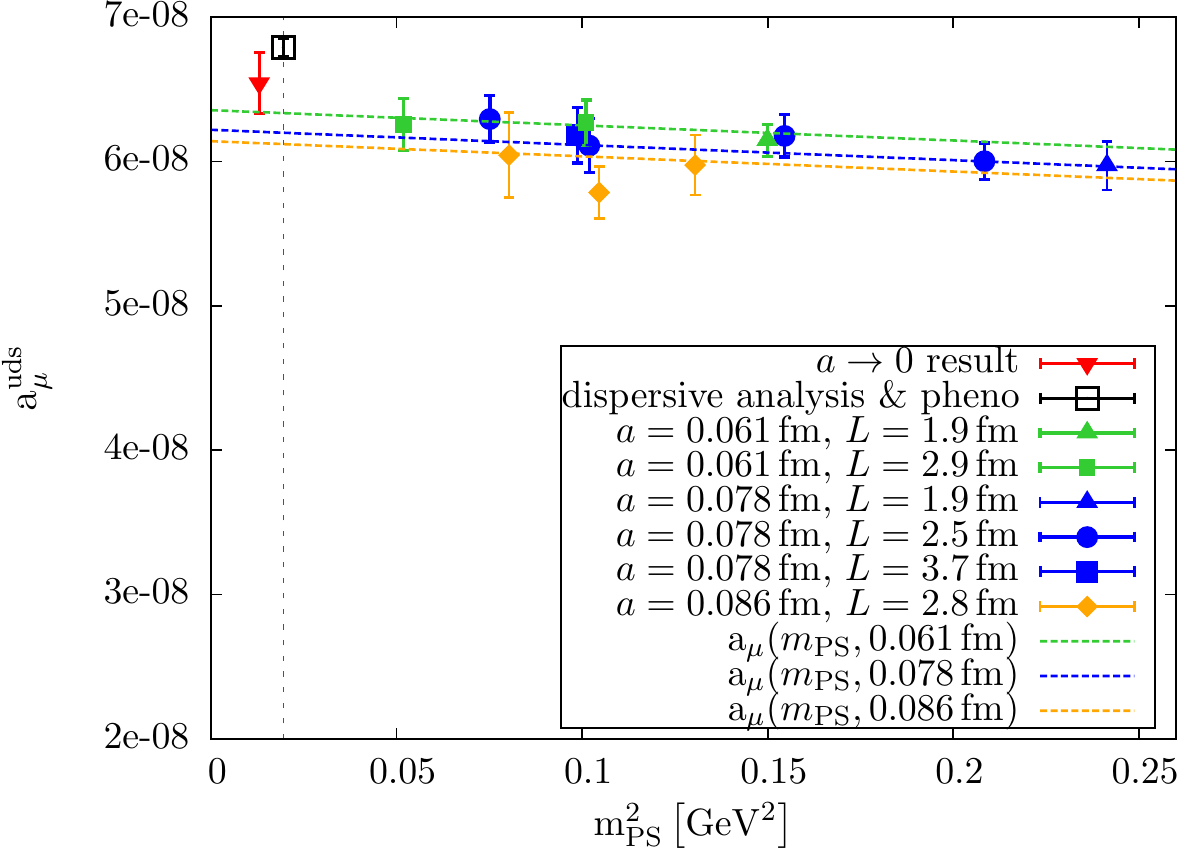}
\caption{Three-flavour contribution to $a_{\mathrm{\mu}}^{\mathrm{hvp}}$. The 
phenomenological value is extracted from~\cite{Jegerlehner:2008zz}
assuming quark-hadron duality.} 
\label{fig:amunf21}
\end{minipage}
\end{figure}

Adding the strange quark in the valence sector, we find that lattice artefacts can no longer be neglected as shown in
Fig.~\ref{fig:amustrange_cl}. Here, only the contribution of the strange quark, $a_{\mathrm{\mu, s}}^{\rm hvp}$, at a fixed pion mass of
about $320\,{\rm MeV}$ is depicted. We have again used our redefinition Eq.~\refeq{eq:amudefresc} with the $\rho$-meson mass as hadronic
scale $H$. This observation is confirmed when performing a combined chiral and continuum extrapolation as we now get a non-zero coefficient
of the $a^2$ term. In the following, we use this combined extrapolation to obtain the results at the physical point in the
continuum limit
\begin{equation}
 a_{\mathrm{\mu}}(m_{\rm PS}, a) = A + B~m_{PS}^2 + C~a^2
\label{eq:fit}
\end{equation}
with $A, B, C$ denoting the free parameters of the fit. 

In Fig.~\ref{fig:amunf21} our three-flavour value obtained in this way in the limit $a\to 0$ is represented by the red triangle slightly
displaced from the physical pion mass to facilitate the comparison with the dispersive result. However, in order to compare the
three-flavour contribution with a
result from a dispersive analysis, we need to disentangle the quark flavours. 
There are 
different possibilities to carry out such a reweighting of the total $a_{\mathrm{\mu}}^{\mathrm{hvp}}$ from a dispersive analysis. We
have reweighted the values
given in~\cite{Jegerlehner:2008zz} with the sum of squared charges of the active flavours assuming quark-hadron duality.
We indicate by the abbreviation ``pheno'' that a certain phenomenological analysis has been 
employed. Comparing our lattice result with this phenomenological extraction method leads to
\begin{eqnarray}
a_{\mathrm{\mu, uds}}^{\rm hvp} & = & 6.55(21)\cdot 10^{-8}\;\;\; (N_f=2+1+1)  \nonumber\\
a_{\mathrm{\mu, uds}}^{\rm hvp} & = & 6.79(05)\cdot 10^{-8}\;\;\; (\mathrm{pheno})
\label{eq:nf21contribution}
\end{eqnarray}
where we find, at least within the errors, an agreement. Our three-flavour result is also compatible with the other lattice computations of
Refs.~\cite{Boyle:2011hu}.


\subsection{The four-flavour contribution, $a_{\mathrm{\mu}}^{\rm hvp}$}
\begin{figure}[htb]
\begin{minipage}{0.53 \textwidth}
\includegraphics[width=\textwidth]{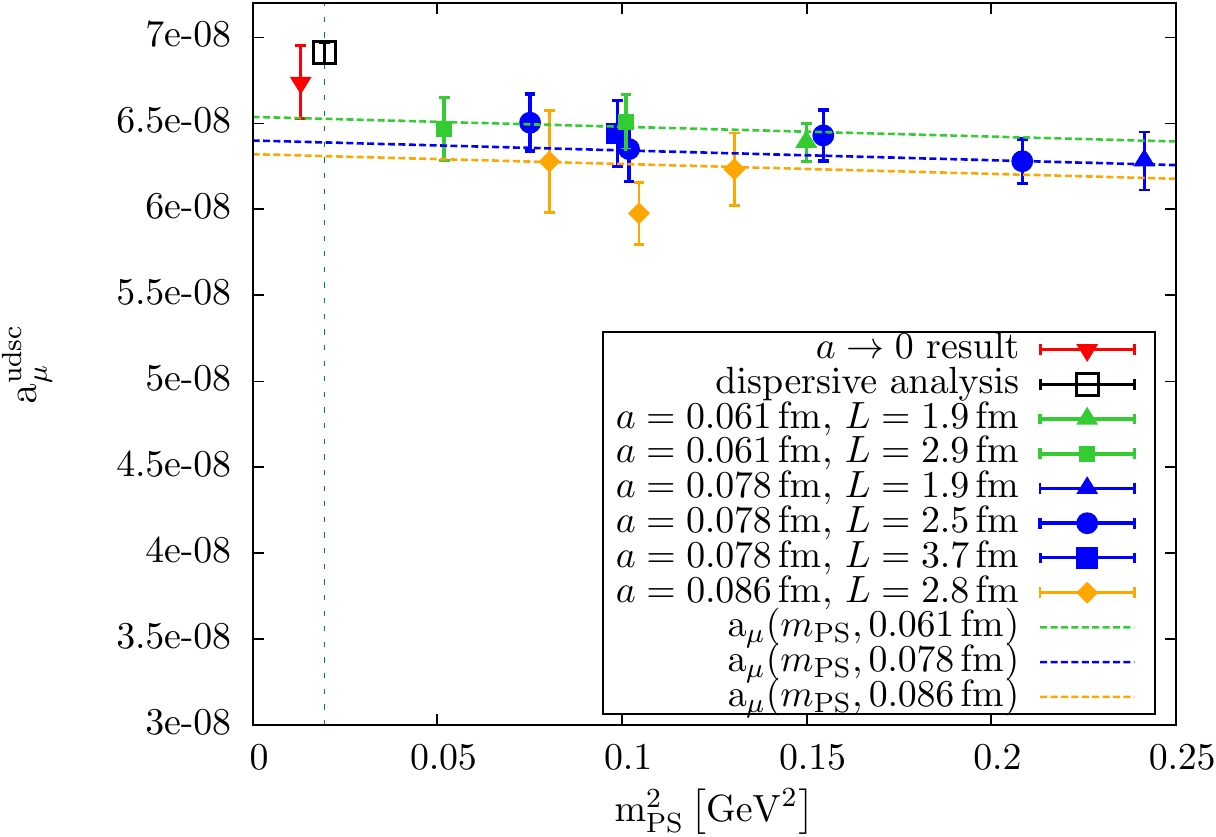}
\caption{$N_f=2+1+1$ result for $a_{\mathrm{\mu}}^{\rm hvp}$.} 
\label{fig:amutot}
\end{minipage}
\hspace{0.02\textwidth}
\begin{minipage}{0.4\textwidth}
\includegraphics[width=\textwidth]{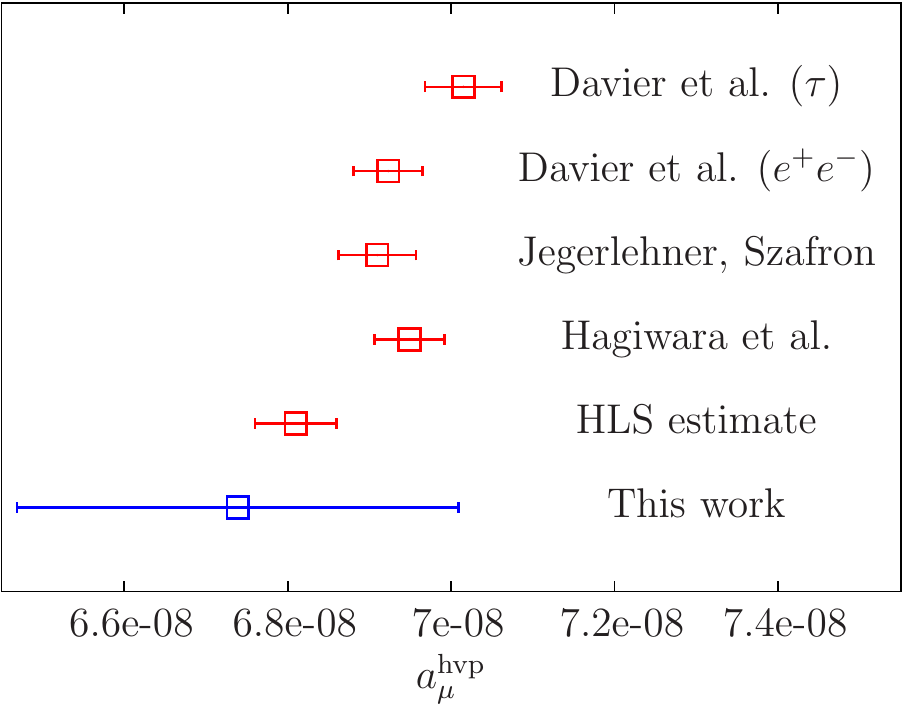}
\caption{Comparison of our first four-flavour lattice result
of $a_{\mathrm{\mu}}^{\rm hvp}$ with different results based on dispersion relations:
Davier et al.~\cite{Davier:2010nc},
Jegerlehner and Szafron~\cite{Jegerlehner:2011ti},
Hagiwara et al.~\cite{Hagiwara:2011af}, and HLS~\cite{Benayoun:2012wc}}
\label{fig:amutot_comp}
\end{minipage}
\end{figure}

Incorporating the charm quark contribution according to Eq.~\refeq{eq:amudefresc} again using $H=m_V$  
we are able to
directly compare to experimental values and those from different dispersive analyses without the ambiguity of any additional
phenomenological analysis. 
Since the charm quark is even heavier than the strange quark, we 
again use a combined fit 
of the form stated in Eq.~\refeq{eq:fit}. In this way, we arrive at the picture shown in 
Fig.~\ref{fig:amutot}. Here, our result obtained in the continuum limit
and at the physical value of the pion mass, represented by the red triangle,
can now be unambiguously confronted with the corresponding one from 
a dispersive analysis~\cite{Jegerlehner:2008zz}:
\begin{eqnarray}
a_{\mathrm{\mu}}^{\rm hvp} & = & 6.74(21)(18)\cdot 10^{-8}\;\;\; (N_f=2+1+1)  \nonumber\\
a_{\mathrm{\mu}}^{\rm hvp} & = & 6.91(01)(05)\cdot 10^{-8}\;\;\; (\mathrm{dispersive}\; \mathrm{analysis})\;. 
\end{eqnarray}
The first uncertainty is of statistical nature whereas the second one is the systematic uncertainty, in our case from the choice of fit
function and excited state contamination in the correlator fits. All other systematic effects investigated in~\cite{Burger:2013jya} have
been found to be negligible. Now a convincing agreement between the two ways of determining this important quantity is found. The value of
the total $a_{\mathrm{\mu}}^{\rm hvp}$ can also be compared with the outcome of other calculations utilising a dispersion relation as shown
in Fig.~\ref{fig:amutot_comp}. Due to the larger uncertainty of the lattice calculation it is not yet possible to discriminate between
the various phenomenological results.

\section{Summary and Outlook}
In this proceeding contribution we have reported on the first four-flavour determination of the leading-order hadronic
contribution to the muon anomalous magnetic moment including also for the first time the continuum limit of this quantity. The result
agrees with various results employing the dispersion relation. However, the uncertainty of the lattice calculation is still about five
times bigger
than those of the dispersive analyses. This uncertainty includes an estimate of systematic effects. In particular, we have checked the
chiral extrapolation of the light quark contribution by performing the computation on ETMC's new $N_f=2$ ensemble directly at the physical
point.

In order to be competitive with the phenomenological determinations of this fundamental quantity, we will have to improve the
accuracy of our calculation. To this end, all-mode-averaging introduced in~\cite{Blum:2012uh} looks like a promising method. We also
plan to repeat the calculation on $N_f=2+1+1$ twisted mass ensembles featuring the physical value of the pion mass.

\section*{Acknowledgements}
We thank Andreas Ammon for providing us with the information of 
the matching K- and D-meson 
masses in the mixed-action setup with their physical values.
This work has been supported in part by the DFG Corroborative
Research Center SFB/TR9.
G.H.~gratefully acknowledges the support of the German Academic National Foundation (Studienstiftung des deutschen Volkes e.V.) and of the
DFG-funded Graduate School GK 1504.
K.J. was supported in part by the Cyprus Research Promotion
Foundation under contract $\Pi$PO$\Sigma$E$\Lambda$KY$\Sigma$H/EM$\Pi$EIPO$\Sigma$/0311/16.
This manuscript has been coauthored by Jefferson Science Associates, LLC under Contract No.~DE-AC05-06OR23177 with the U.S.~Department of
Energy.
The numerical computations have been performed on the
{\it SGI system HLRN-II} at the {HLRN Supercomputing Service Berlin-Hannover},  FZJ/GCS, BG/P, and BG/Q at FZ-J\"ulich.


\begin{thebibliography}{99}
\bibitem{Bennett:2006fi}
  G.~W.~Bennett {\it et al.}  [Muon G-2 Collaboration],
  Phys.\ Rev.\ D {\bf 73} (2006) 072003
  [hep-ex/0602035].

\bibitem{Roberts:2010cj}
  B.~L.~Roberts,
  Chin.\ Phys.\ C {\bf 34} (2010) 741
  [arXiv:1001.2898 [hep-ex]].

\bibitem{Davier:2010nc}
  M.~Davier, A.~Hoecker, B.~Malaescu and Z.~Zhang,
  Eur.\ Phys.\ J.\ C {\bf 71} (2011) 1515
   [Erratum-ibid.\ C {\bf 72} (2012) 1874]
  [arXiv:1010.4180 [hep-ph]].

\bibitem{Benayoun:2012wc}
  M.~Benayoun, P.~David, L.~DelBuono and F.~Jegerlehner,
  Eur.\ Phys.\ J.\ C {\bf 73} (2013) 2453
  [arXiv:1210.7184 [hep-ph]].

\bibitem{Blum:2002ii}
  T.~Blum,
  Phys.\ Rev.\ Lett.\  {\bf 91} (2003) 052001
  [hep-lat/0212018].

\bibitem{Aubin:2006xv}
  C.~Aubin and T.~Blum,
  Phys.\ Rev.\ D {\bf 75} (2007) 114502
  [hep-lat/0608011].

\bibitem{Feng:2011zk}
  X.~Feng, K.~Jansen, M.~Petschlies and D.~B.~Renner,
  Phys.\ Rev.\ Lett.\  {\bf 107} (2011) 081802
  [arXiv:1103.4818 [hep-lat]].

\bibitem{Boyle:2011hu}
  P.~Boyle, L.~Del Debbio, E.~Kerrane and J.~Zanotti,
  Phys.\ Rev.\ D {\bf 85} (2012) 074504
  [arXiv:1107.1497 [hep-lat]];
  M.~Della Morte, B.~J\"ager, A.~J\"uttner and H.~Wittig,
   JHEP {\bf 1203} (2012) 055
   [arXiv:1112.2894 [hep-lat]].


\bibitem{Baron:2010bv}
  R.~Baron {\it et al.} [European Twisted Mass Collaboration],
  JHEP {\bf 1006} (2010) 111
  [arXiv:1004.5284 [hep-lat]]; 
  Comput.\ Phys.\ Commun.\  {\bf 182} (2011) 299
  [arXiv:1005.2042 [hep-lat]].


\bibitem{Feng:2012gh}
  X.~Feng, G.~Hotzel, K.~Jansen, M.~Petschlies and D.~B.~Renner,
  PoS LATTICE {\bf 2012} (2012) 174
  [arXiv:1211.0828 [hep-lat]].
  \bibitem{bartek:2013}
   B.~Kostrzewa {\it et al.} [European Twisted Mass Collaboration],
  PoS LATTICE {\bf 2013} (2013) 264.

\bibitem{Burger:2013jya}
  F.~Burger, X.~Feng, G.~Hotzel, K.~Jansen, M.~Petschlies and D.~B.~Renner,
  arXiv:1308.4327 [hep-lat].

\bibitem{Aubin:2012me}
  C.~Aubin, T.~Blum, M.~Golterman and S.~Peris,
  Phys.\ Rev.\ D {\bf 86} (2012) 054509
  [arXiv:1205.3695 [hep-lat]].

\bibitem{Golterman:2013vna}
  M.~Golterman, K.~Maltman and S.~Peris,
  arXiv:1309.2153 [hep-lat]; arXiv:1310.5928 [hep-lat].


\bibitem{Jegerlehner:2008zz}
  F.~Jegerlehner,
  Nucl.\ Phys.\ Proc.\ Suppl.\  {\bf 181-182} (2008) 26.

\bibitem{Jegerlehner:2011ti}
  F.~Jegerlehner and R.~Szafron,
  Eur.\ Phys.\ J.\ C {\bf 71} (2011) 1632
  [arXiv:1101.2872 [hep-ph]].

\bibitem{Hagiwara:2011af}
  K.~Hagiwara, R.~Liao, A.~D.~Martin, D.~Nomura and T.~Teubner,
  J.\ Phys.\ G {\bf 38} (2011) 085003
  [arXiv:1105.3149 [hep-ph]].

\bibitem{Blum:2012uh}
  T.~Blum, T.~Izubuchi and E.~Shintani,
  arXiv:1208.4349 [hep-lat].

\end{thebibliography}
\end{document}